\documentclass[
aps,
prl,
showpacs
amsmath,
amssymb,
floatfix,
longbibliography,
letterpaper,
lengthcheck,
superscriptaddress,
nofootinbib
]{revtex4-1}

\usepackage{graphicx}
\usepackage{rotating}
\usepackage{amsmath}
\usepackage{bbm}
\usepackage{subfigure}
\usepackage{color}
\usepackage{braket}
\begin{document}

\title{Atoms and Molecules in Cavities: From Weak to Strong Coupling in QED Chemistry}

\author{Johannes Flick}
  \email[Electronic address:\;]{johannes.flick@mpsd.mpg.de}
  \affiliation{Max Planck Institute for the Structure and Dynamics of Matter and Center for Free-Electron Laser Science Department of Physics, Luruper Chaussee 149, 22761 Hamburg, Germany}
  \affiliation{Fritz-Haber-Institut der Max-Planck-Gesellschaft, Faradayweg 4-6, D-14195 Berlin-Dahlem, Germany}
\author{Michael Ruggenthaler}
  \email[Electronic address:\;]{michael.ruggenthaler@mpsd.mpg.de}
  \affiliation{Max Planck Institute for the Structure and Dynamics of Matter and Center for Free-Electron Laser Science Department of Physics, Luruper Chaussee 149, 22761 Hamburg, Germany}
\author{Heiko Appel}
  \email[Electronic address:\;]{heiko.appel@mpsd.mpg.de}
  \affiliation{Max Planck Institute for the Structure and Dynamics of Matter and Center for Free-Electron Laser Science Department of Physics, Luruper Chaussee 149, 22761 Hamburg, Germany}
  \affiliation{Fritz-Haber-Institut der Max-Planck-Gesellschaft, Faradayweg 4-6, D-14195 Berlin-Dahlem, Germany}
\author{Angel Rubio}
  \email[Electronic address:\;]{angel.rubio@mpsd.mpg.de}
  \affiliation{Max Planck Institute for the Structure and Dynamics of Matter and Center for Free-Electron Laser Science Department of Physics, Luruper Chaussee 149, 22761 Hamburg, Germany}
  \affiliation{Fritz-Haber-Institut der Max-Planck-Gesellschaft, Faradayweg 4-6, D-14195 Berlin-Dahlem, Germany}
  \affiliation{Nano-Bio Spectroscopy Group and ETSF,  Dpto. Fisica de Materiales, Universidad del Pa\'is Vasco, 20018 San Sebasti\'an, Spain}

\date{\today}

%%%%%%%%%%%%%%%%%%%%%%%%%%%%%%%%%%%%%%%%%%%%%%%%%%%%%%%%%%%%%%%%%%
%                            Abstract                            %
%%%%%%%%%%%%%%%%%%%%%%%%%%%%%%%%%%%%%%%%%%%%%%%%%%%%%%%%%%%%%%%%%%
\begin{abstract}
In this work, we provide an overview of how well-established concepts in the fields of quantum chemistry and material sciences have to be adapted when the quantum nature of light becomes important in correlated matter-photon problems. Therefore, we analyze model systems in optical cavities, where the matter-photon interaction is considered from the weak- to the strong coupling limit and for individual photon modes as well as for the multi-mode case. We identify fundamental changes in Born-Oppenheimer surfaces, spectroscopic quantities, conical intersections and efficiency for quantum control. We conclude by applying our novel recently developed quantum-electrodynamical density-functional theory~\cite{ruggenthaler2014,flick2015} to single-photon emission and show how a straightforward approximation accurately describes the correlated electron-photon dynamics. This paves the road to describe matter-photon interactions from first-principles and addresses the emergence of new states of matter in chemistry and material science.
\end{abstract}

\pacs{71.15.-m, 31.70.Hq, 31.15.ee}
% 71.15.-m Methods of electronic structure calculations
% 31.70.Hq Time-dependent phenomena: excitation and relaxation
% 31.15.ee Time-dependent density functional theory

\date{\today}

\maketitle

%%%%%%%%%%%%%%%%%%%%%%%%%%%%%%%%%%%%%%%%%%%%%%%%%%%%%%%%%%%%%%%%%%
%                        Introduction                            %
%%%%%%%%%%%%%%%%%%%%%%%%%%%%%%%%%%%%%%%%%%%%%%%%%%%%%%%%%%%%%%%%%%
{N}ovel experimental possibilities have allowed scientists to obtain new insights into how photons interact with matter and how these interactions correlate photonic and particle degrees of freedom. Such experiments show, e.g., an increase of the conductivity in organic semiconductors through hybridization with the vacuum field~\cite{orgiu2015}, strong shifts of the vibrational frequencies by the coupling of molecular resonators with a microcavity mode~\cite{shalabney2015}, non-classical single photon-phonon correlations~\cite{riedinger2016}, the control of spin relaxations using an optical cavity~\cite{bienfait2016}, the enhancement of Raman scattering from vibro-polariton 
states~\cite{shalabney2015a,pino2015}, single molecule strong coupling~\cite{chikkaraddy2016}, sampling of vacuum fluctuations~\cite{riek2015}, strong exciton-photon coupling of light-harvesting complexes~\cite{coles2014}, strong long-range atom-atom interactions mediated by photons~\cite{gonzales2015}, attractive photonic states~\cite{firstenberg2013,maghrebi2015}, or superradiance for atoms in photonic crystals~\cite{goban2015}. All these results indicate the appearence of new states of matter and subsequently a change in the chemical properties of the matter system~\cite{feist2015,schachenmayer2015,cirio2016,kowalewski2016}, if the quantum nature of light becomes important. This is the case, e.g., in so-called strong-coupling situations, which are nowadays of central interest in the fields of circuit quantum electrodynamics (circuit-QED)~\cite{wallraff2004,blais2004,rossatto2016} or cavity-QED~\cite{benson2011,ritsch2013}. While the analysis of such experiments are routinely performed with the help of simplified (few-level) models that are able to capture the essential physics, for the (quantitative) prediction of properties of complex multi-particle systems coupled to photons, methods that can treat such coupled boson-fermion situations from first principles seem worthwhile~\cite{ruggenthaler2011b, tokatly2013, ruggenthaler2014, sakkinen2014, sakkinen2015, flick2015, melo2016}. On the other hand, the strong coupling to photons can challenge our conventional understanding of electronic structures and allows to study the influence of the quantum nature of light on chemical processes.
\\
In this work we want to highlight the possibilities as well as the theoretical challenges that arise at the interface of electronic structure theory and quantum optics. To this end we discuss three distinct situations where the photon-matter correlation becomes significant and modifies conventional concepts of electronic-structure theory and quantum optics. In part (\textbf{1}) we study systems, which contain nuclear, electronic and photonic degrees of freedom explicitly. First we consider a model dimer molecule that contains two nuclei and two electrons confined to one dimension and which is placed in an optical high-Q cavity. We show how the photons change the electronic Born-Oppenheimer (BO) surfaces in a complex way. These changes affect, e.g., the bond-length and the absorption spectrum of the molecule~\cite{galego2015}. Additionally, we show how the ground-state of the full system obtains an electron-nuclear(vibronic)-photon quasiparticle character, the vibro-polariton. The second model system, we study in part (\textbf{1}), is the two-dimensional Shin-Metiu model~\cite{shin1996,min2014}, which consists of three nuclei and a single electron located in an optical high-Q cavity in resonance to the lowest vibrational excitation. The two-dimensional Shin-Metiu, which is a model system for an $H_3$ molecule featuring a conical intersection in the Born-Oppenheimer surfaces, and we show how this intersection can be altered in the case of strong light-matter interactions. In part (\textbf{2}) we show how the control of electronic systems~\cite{castro2014} is modified if we take into account the coupling to a cavity mode. These extra degrees of freedom allow to achieve a predefined target more efficiently with less external driving, when either the cavity frequency or the electron-photon coupling are chosen in favor. Additionally the external driving of the photonic field by external dipoles allows to influence the electron transport to gain efficiency. In the last part of this paper, part (\textbf{3}), we consider single-photon emission and how photon-bound polariton states appear in multi-mode cavities, if the matter-photon coupling is increased to strong coupling. This leads, e.g., to the breakdown of the Purcell effect~\cite{liberato2014}. For such a strong-coupling situation we demonstrate the capabilities of the recently developed density-functional theory for cavity QED systems~\cite{tokatly2013, ruggenthaler2014, pellegrini2015,flick2015}. We show the limitations of a semi-classical treatment and that the first approximate exchange-correlation functional~\cite{pellegrini2015} for cavity-QED systems along the line of the optimized-effective potential (OEP) approach~\cite{kuemmel2003} allows to accurately treat such situations.

\section{Cavity QED - The molecular dimer case}

Let us start with a model that contains all the major degrees of freedom of a real system: nuclear, electronic and photonic. The model we consider is an artificial one-dimensional molecule that consists of two nuclei and two electrons. In a traditional quantum-chemical treatment the photonic degrees of freedom would be neglected since one assumes the multi-particle system to be in free space and the Coulomb interaction\footnote{The Coulomb interaction can be inferred from QED~\cite{greiner1996, ruggenthaler2014}, where the longitudinal part of the photon field is solved explicitly in terms of the longitudinal charge current of the particles. Thus, this assumption seems well-justified whenever the transversal currents of the particle system are negligible.} is supposed to describe the major contribution of the interaction due to the photon field. However, if we put the molecule inside an optical cavity, we change the photon modes\footnote{We point out that this also changes the interaction due to the longitudinal currents and hence the Coulomb interaction is modified. However, this effect is beyond the scope of this work and will be explored in a separate publication.} and find situations where the photon degrees of freedom play a crucial role. To investigate this situation, we consider as first example a molecule inside a cavity where one of the modes is tuned to the first vibrational excitation of the dimer system. This is the photon degree of freedom that we will keep in our calculations. We show how this can affect standard concepts of electronic-structure theory, e.g., the BO surfaces. Schematically, this electron-nuclear-photon system can be understood as follows:
\begin{figure}[h] 
 \centerline{\includegraphics[width=0.5\textwidth]{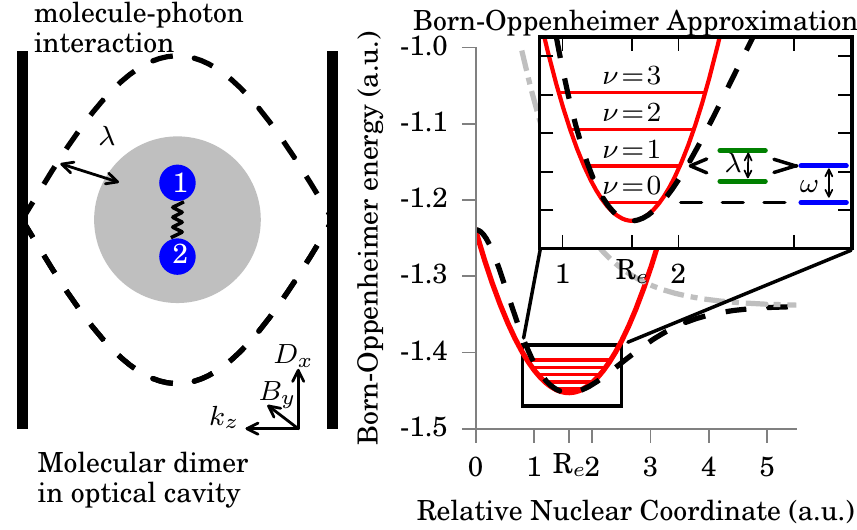}}
 \caption{Left: Schematic illustration of the dimer in a cavity coupled to a single-mode field polarized along the $x$ axis. Right: Exact ground-state BO surface in black dashed lines, harmonic ground-state BO surface in solid red lines, and exact first-excited-state BO surface in dashed-dotted grey lines; $\nu$ indicates the phonon excitation, $R_e$ marks the BO equilibrium distance, and $\lambda$ denotes the Rabi splitting by the phonon-photon hybridization and the matter-photon interaction strength. The photon wavevector $k_z$, the magnetic field $B_y$, and the displacement field $D_x$ build a triad~\cite{craig1998}.}
\label{fig:hd-dimer}
\end{figure}

\begin{figure}[h] 
\centerline{\includegraphics[width=0.5\textwidth]{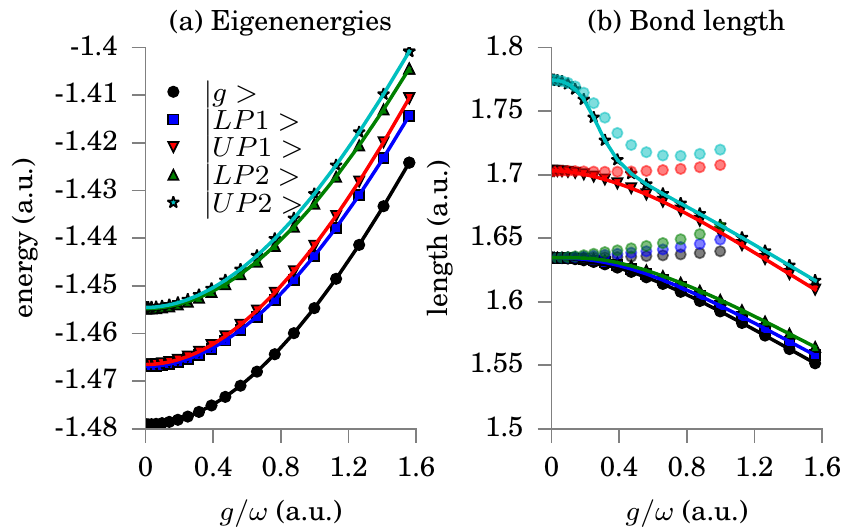}}
\caption{(a) shows the eigenenergies for different values of the matter-photon interaction strength $\lambda$, (b) shows the bond length $\left<\textbf{X} \right>$. In both plots, black dots correspond to the ground-state $\ket{g}$, blue squares to the lower polariton state $\ket{LP}$ and red lower triangle to the upper polariton state $\ket{UP}$, green upper triangles to the second lower polariton state $\ket{LP2}$ and stars in cyan plots to the second upper polariton state $\ket{UP2}$. In (b) we denote the bond-length values by shaded colored dots without considering the $R^2$ term of Eq.~[\ref{eq:qed-chemistry-hamiltonian}].}\label{fig:qed-chemistry-results1}
\end{figure}

On the left of Fig.~\ref{fig:hd-dimer} the molecule in the cavity is shown. The molecule is exposed to a single cavity mode, which is given by one of the cavity frequencies $\omega_\alpha$ and the matter-photon coupling strength $\lambda_\alpha$. On the right side of Fig.~\ref{fig:hd-dimer} we depict a simplified picture of the hybridization of the system. We show the BO surface depending on the nuclear coordinate ${X}$ in atomic units (Bohr) and indicate the eigenstates of the molecular system in BO approximation. In the ground state, the electrons are subject to the ground-state BO surface, which is shown in dashed-black lines. The harmonic approximation to this full surface is shown in solid red lines. The individual harmonic excitations of the nuclear (phonon) subsystem are indicated by the quantum number $\nu$. Since the cavity mode is tuned in resonance, we find Rabi-splitting~\cite{shore1993} of the first vibrational excitation, which is proportional to the matter-photon coupling constant ${\lambda_\alpha}$. The first excited electronic BO surface is shown in gray dashed lines. This surface has no minima, hence featuring the dissociation of the molecule. In the dissociation limit ($X>5$ a.u.), the ground-state and the first-excited BO surfaces merge. 
\\
For a detailed investigation in the following, we consider this system in the dipole approximation and in the length gauge. In this setup, the general correlated electron-nuclear-photon Hamiltonian consisting of $n_e$ electrons, $n_n$ nuclei, and $n_{p}$ photon modes can be written as a sum of the electro-nuclear Hamiltonian $\hat{H}_{en}$ and the photon Hamiltonian $\hat{H}_p$~\cite{faisal1987,tokatly2013,pellegrini2015,flick2015}:
\begin{align}
\label{eq:qed-chemistry-hamiltonian}
\hat{H} &= \hat{H}_{en} + \hat{H}_p\\ 
\hat{H}_{en} &= \hat{T}_e + \hat{T}_N + \hat{W}_{ee} + \hat{W}_{NN} + \hat{W}_{eN}\\
\hat{H}_p &= \frac{1}{2}\sum\limits_{\alpha=1}^{n_p}\left[\hat{p}^2_\alpha+\omega_\alpha^2\left(\hat{q}_\alpha + \frac{\boldsymbol \lambda_\alpha} {\omega_\alpha} \cdot{e\textbf{R}} \right)^2\right]\label{eq:ham-photons}\\
\label{eq:dipole_operator}
{\textbf{R}} & = \sum_{I=1}^{N_n} Z_I\textbf{X}_I  - \sum_{i=1}^{N_e} \textbf{x}_i,
\end{align} 
where $Z_I$ specifies the nuclear charges. The kinetic energy is given by $\hat{T} = \sum_{i=1} {\hbar^2}\vec{\nabla}^2_{\textbf{x}_{i}}/{2m_i}$, for electrons and nuclei with mass $m_i$, respectively. Further, instead of the bare Coulomb interaction we use a soft-Coulomb interaction~\cite{loudon1959} for $\hat{W}$ as routinely done for one-dimensional model systems, i.e. $\hat{W} = \sum_{i,j>i} Z_i Z_j /4\pi\epsilon_0 \sqrt{\left(\textbf{x}_i-\textbf{x}_j\right)^2+1}$, with resulting negative (positive) prefactor for the electron-nuclear (electron-electron/nuclear-nuclear) interaction. In the following, for the specific dimer systems, the capital variables, $\textbf{X}_1$ and $\textbf{X}_2$, denote the nuclear coordinates, while the small variables $\textbf{x}_3$ and $\textbf{x}_4$ denote the electronic coordinates, and $\hat{q}_\alpha=\sqrt{\frac{\hbar}{2\omega_\alpha}}\left(\hat{a}_\alpha^\dagger + \hat{a}_\alpha \right)$ defines the photon displacement coordinate using the photonic creation and annihilation operators~\cite{tokatly2013,pellegrini2015}. The photon displacement operator is connected to the electric displacement field operator {$\hat{\textbf{D}}_\alpha = {\omega_\alpha\boldsymbol\lambda_\alpha} \hat{q}_\alpha$}, where ${\boldsymbol\lambda_\alpha}$ is the transversal polarization vector times the dipole-approximation coupling strength $\lambda_\alpha$. Using $g_\alpha = \sqrt{\frac{\hbar\omega_\alpha}{2}}{\lambda_\alpha}$, we can connect to typical strong-coupling calculations as, e.g., in reference~\cite{galego2015}. We only describe the two valence electrons explicitly. To this end, we choose for the nuclear masses $M_1=m_p$ and $M_2=m_p$, where $m_p$ is the proton mass and with nuclear charges $Z_1 = 1.2$, and $Z_2=0.8$. The electron masses correspond to the electron mass $m_e$, i.e. $m_3=m_4=m_e$.  In the photon Hamiltonian $\hat{H}_p$, we consider the electron-nuclear-photon coupling in dipole approximation, where $\textbf{R}$ is full dipole operator that contains both the electronic and nuclear contributions. The complete many-body problem including two electrons, two nuclei and one photon mode is a five-dimensional problem. To reduce the computational complexity, we perform a coordinate transformation into a center-of-mass frame such that the center-of-mass motion can be separated and we are left with a four dimensional problem for the internal degrees of freedom~\cite{crawford2015}. For details on the transformation and the real-space grid used to perform the numerical calculation, we refer the reader to (SI1). For clarity, we will use the original Euclidean coordinates in all formulas throughout this paper with the exception of the nuclear relative coordinate $\textbf{X} = \textbf{X}_1-\textbf{X}_2$. The cavity frequency $\omega_\alpha$ is chosen to be in resonance to the first vibrionic transition $\omega_{12}$, hence $\omega_\alpha = \omega_{12} = 0.01216 \text{a.u.}$. The dipole moment of this transition has a value of $d_{12}=0.01869 \text{a.u.}$.\\
The Hamiltonian in Eq.~(\ref{eq:qed-chemistry-hamiltonian}) contains besides the (softened) Coulomb interactions two new interaction terms: the explicit dipolar matter-photon coupling $ \sum_\alpha\omega_\alpha\hat{q}_\alpha\left({\boldsymbol \lambda_\alpha\cdot e\textbf{R}}\right)$ and the quadratic dipole self-energy term $\sum_\alpha{\left({\boldsymbol\lambda_\alpha}\cdot e\textbf{R}\right)^2}/{2}$. The dipole self-energy term is the analogue of the $A^2$ term in the momentum-gauge, i.e., represents how the electrons act back and change the frequency and polarization of the photon field. This term is usually neglected and only rarely considered~\cite{faisal1987,todorov2010,todorov2012,bamba2015}. It is a clear relevant beyond two-levels effect, since in that case the dipole self-energy term reduces to a constant energy offset in the case of a two-level approximation, such as the Jaynes-Cummings-Model~\cite{shore1993,pellegrini2015}. {Additionally recent experiments have arrived at the same conclusion, i.e. the particular importance of such a dipole self-energy term in the strong-coupling regime~\cite{george2016}}. Furthermore, in an unconfined cavity-free three-dimensional setup, it is usually neglected in the inter-molecular region, where it cancels the inter-molecular Coulomb interaction~\cite{craig1998,vukics2014,vukics2015}, or in the limit of dilute atomic gases and infinite quantization volume~\cite{faisal1987}. However, in the intra-molecular region in a cavity, which is the focus of the present study, this term has to be taken into account as becomes obvious from Fig.~\ref{fig:qed-chemistry-results1}.
%Additionally, we also want to point out that in Eq.~(\ref{eq:qed-chemistry-hamiltonian}), the Coulomb interaction is considered in the cavity-free limit. We also expect a slightly different form of the Coulomb interaction for cavity systems, in particular in the strong-coupling regime. However this effect is beyond the scope of this paper and will be explored in a separate publication.\\
In Fig.~\ref{fig:qed-chemistry-results1} (a) we show the exact eigenenergies of the cavity system obtained by exact diagononalization~\cite{flick2014} as function of the matter-photon coupling strength $ \lambda_\alpha$. The general harmonic trend is given by the self-polarization interaction term. In black, we plot the ground-state energy, in red/blue (cyan/green), we plot the first (second) upper and lower polariton states. The matter-photon coupling induces the Rabi-splitting in the energy, as illustrated in Fig.~\ref{fig:hd-dimer}. With increasing $ \lambda_\alpha$, we find an increasing Rabi splitting. The bond-length $\left<\textbf{X}\right>$ of the individual states is plotted in Fig.~\ref{fig:qed-chemistry-results1} (b). For this plot the same color code as in (a) applies and additionally, we plot the bond-length values in shaded colors of the states, if we neglect the dipole self-energy term in Eq.~[\ref{eq:qed-chemistry-hamiltonian}]. We find that the full matter-photon coupling of Eq.~[\ref{eq:qed-chemistry-hamiltonian}] introduces large changes in the bond-length. Here the bond-length is reduced from 1.63 a.u. to 1.55 a.u. by around 5\% for the ground-state. In contrast, if we neglect the dipole self-energy term in Eq.~[\ref{eq:qed-chemistry-hamiltonian}], we find an increasing bond-length with increasing electron-photon coupling and the system is only stable (bound) up to $g/\omega=0.9$. This finding is a clear indication of the importance of the usually neglected dipole self-energy term in the strong-coupling limit and agrees with recent experimental findings~\cite{george2016}.
\begin{figure}[h]  
\centerline{\includegraphics[width=0.5\textwidth]{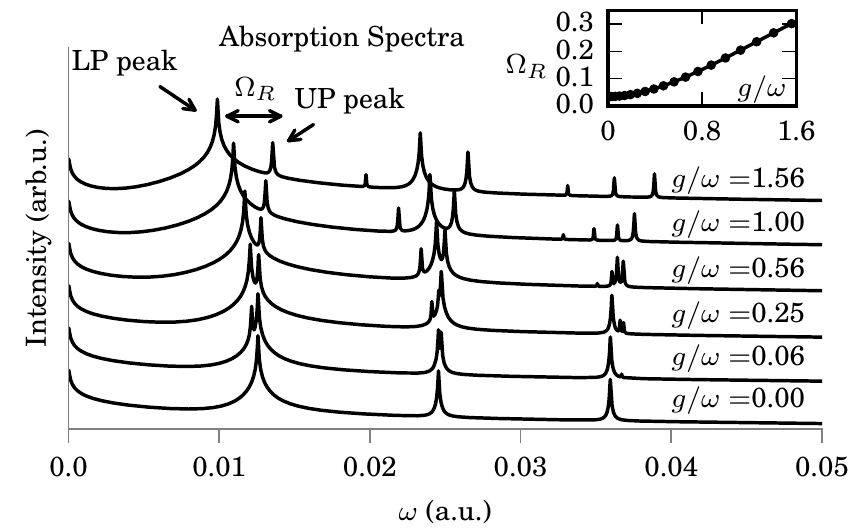}}
\caption{Calculated absorption spectra for the dimer in a cavity of Fig.~\ref{fig:hd-dimer} for different values of the matter-photon coupling strength $g/\omega$. The first two peaks correspond to the lower polariton peak (LP) and upper polariton peak (UP). In the inset, we quantify the Rabi splitting $\Omega_R$ as function of the coupling constant. (see text for details.)}\label{fig:qed-chemistry-results3}
\end{figure}
Next, we show how a spectroscopic quantity of our electron-nuclear-photon model is influenced by strong matter-photon coupling. To this end we determine the ground-state absorption spectrum using a sum-over-states expression~\cite{galego2015} explained in (SI4). In Fig.~\ref{fig:qed-chemistry-results3} we show spectra for different values of the matter-photon coupling strength ${\lambda}_\alpha$. For increasing coupling, we find clear signatures of a strong Rabi-splitting $\Omega_R=\left(E_3-E_2\right)/\omega_\alpha$, where $E_3$ and $E_2$ are the eigenvalues of Eq.~[\ref{eq:qed-chemistry-hamiltonian}]. In the spectra, we explicitly denote the lower polaritonic and the upper polaritonic peak, which become clearly visible in the strong-coupling limit. Additionally, higher lying excitations also show Rabi-splitting, e.g., the second peak shows a three-fold splitting. In the inset of the figure, we show that in the range of the used parameters, the Rabi-splitting goes up to $0.3$. For the matter-photon coupling strength, we choose values between $0 \leq g\leq 1.6 \omega_\alpha$. Recent experiments, as e.g. Refs.~\cite{shalabney2015,george2016}, report Rabi-splittings from $0.1-0.25$ and as seen in the inset a value of $g=1.6\omega$ corresponds to a Rabi splitting of around $0.3$. 

\section{Cavity Born-Oppenheimer (CBO) approximation}

\begin{figure}[h] 
\centerline{\includegraphics[width=0.5\textwidth]{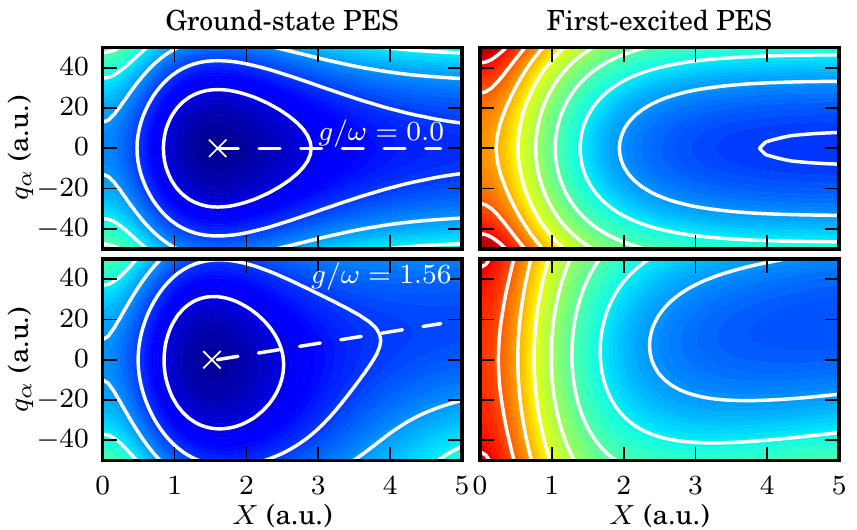}}
\caption{The first row shows the ground state and first-excited two-dimensional CBO surface for $g/\omega=0.0$. Second row the surfaces in the strong-coupling limit for $g/\omega=1.56$. The dashed lines indicate the direction of the normal modes featuring {dissociation} of the system. The $x$ indicates the equilibrium bondlength.}\label{fig:qed-chemistry-results2d}
\end{figure}

\begin{figure}[h] 
\centerline{\includegraphics[width=0.5\textwidth]{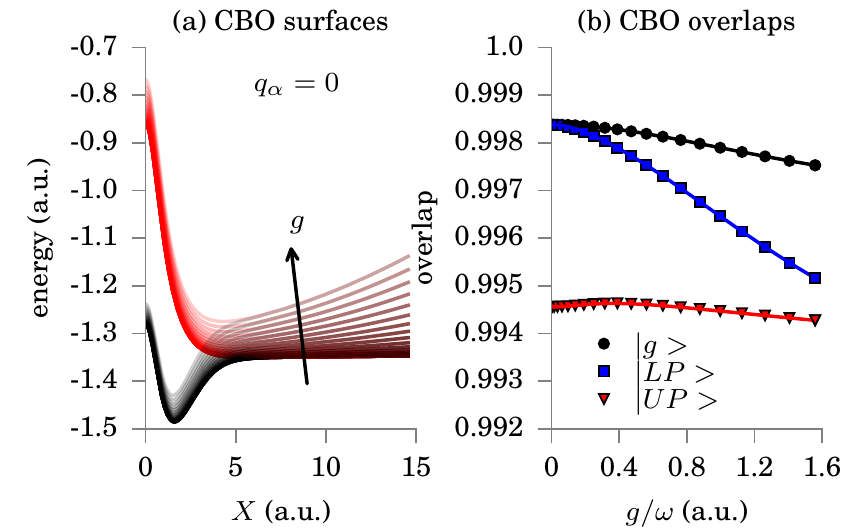}}
\caption{(a) shows the ground state and first-excited CBO surface for different values of the matter-photon interaction strength $\lambda$, (b) shows the overlap of the exact states with the CBO states. Black dots correspond to the ground-state $\ket{g}$, blue dots to the lower polariton state $\ket{LP}$ and red dots to the upper polariton state $\ket{UP}$.}\label{fig:qed-chemistry-results2}
\end{figure}

To highlight the effect that the photons can have on quantum-chemical concepts, we compare the exact calculations done with the above Hamiltonian to a BO calculation that takes the photons into account. This cavity Born-Oppenheimer (CBO) approximation is introduced in (SI2/3).
{ In the CBO approximation, the electronic Hamiltonian $\hat{H}_e\left(\left\{ \textbf{X}\right\},\left\{ \textbf{q}_\alpha\right\}\right)$ parametically depends on all nuclear coordinates $\{\textbf{X}\}$ and photon displacement coordinates $\{\textbf{q}_\alpha\}$. This parameterical dependency is inhereted to the multi-dimensional potential energy surfaces (PES) $V_j\left(\left\{ \textbf{X}\right\},\left\{ \textbf{q}_\alpha\right\}\right)=E_j\left(\left\{ \textbf{X}\right\},\left\{ \textbf{q}_\alpha\right\}\right) + V_{nn}\left(\left\{ \textbf{X}\right\}\right) + \sum_\alpha\omega_\alpha^2q^2_\alpha$, where $E_j$ are the eigenvalues of the electronic CBO Hamiltonian. Such a procedure reduces in the case of ${ \lambda_\alpha} = 0$ to the usual BO approximation~\cite{gross1991}. For more details, we refer the reader to (SI2) and (SI3).} In Fig.~(\ref{fig:qed-chemistry-results2d}) we explicitly show different CBO surfaces. %obtained from the Hamiltonian of Eq.~[\ref{eq:ham-cboa}]. 
These surfaces are two-dimensional surfaces and depend for the dimer system on the nuclear coordinate ${\boldsymbol X}$ and the photon displacement coordinate $q_\alpha$. The left surfaces are the ground-state surfaces, while in the right-side we plot the first-excited state surfaces. In the first row, we plot the surfaces for vanishing matter-photon coupling. Both surfaces show along the x-axis the behavior as in Fig.~\ref{fig:hd-dimer}, while along the y-axis we find a harmonic potential that is associated with the photon coordinate. These surfaces show that we can easily distingish between the photon and nuclear degrees of freedom. In the second row of Fig.~(\ref{fig:qed-chemistry-results2d}) we show the surfaces in the strong-coupling limit. Here, we find that new normal coordinates appear, that are true polaritonic degrees of freedom. The normal coordinates have now photonic and nuclear degrees of freedom.
In Fig.~(\ref{fig:qed-chemistry-results2}) (a) we explicitly show different CBO surfaces in a cut along the photon-coordinate $q_\alpha=0$. {In the figure, in black, we plot the ground state surfaces and in red the first-excited state surfaces.} We find that for increasing $\lambda_\alpha$ the polarization term introduces a harmonic (parabolic) barrier, which alters the BO surfaces significantly.
 The lowest surfaces in the figure corresponds to the cavity-free limit. {This surfaces has a flat tail for large ${\boldsymbol X}$}. 
We see that tuning ${\lambda_\alpha}$ allows to shape the BO surfaces harmonically {that has in particular implications on the tail of the surface}. In general, changes in the BO surfaces alter the chemistry of the system, with implications on various quantities, e.g., the bond length, tunneling barriers, or transition rates. For instance, since the nuclear coordinate ${\boldsymbol X}$ in Fig.~(\ref{fig:qed-chemistry-results2}) is also a measure for the nuclear bond length, we find that increasing the value of ${\lambda_\alpha}$ shifts the bond length to smaller values. As in the exact calculation, the opposite trend would be found if we neglected the polarization contribution to the matter-photon coupling. In red, we show the first-excited state surfaces. While these surfaces feature the dissociation of the molecule in the cavity-free case, we find a local minimum of the surfaces for strong matter-photon coupling $\lambda_\alpha$. We emphasize, however, that along the new normal-coordinate direction indicated in Fig.~\ref{fig:qed-chemistry-results2d} the system is dissociating, as in the field free case for $q_\alpha=0$. Since in the strong-coupling limit the photonic and nuclear degrees of freedom are highly correlated the dissociation also corresponds to an excitation in the photonic degree of freedom. Next we assess the quality of the CBO approximation and plot the overlap of the cavity-BO approximated wave functions with the exact wave functions in Fig.~\ref{fig:qed-chemistry-results2} (b). While for small values of $\lambda_\alpha$ the CBO approximation has the same quality as the cavity-free electronic BO approximation, we observe lower overlaps for strong matter-photon interaction. The overlap of the CBO upper polariton state with the exact correlated state drops to 0.994\% in the strong-coupling limit. Following the usual trend known from the standard BO approximation, the quality of higher-lying states, which here are the upper and the lower polariton states, is lower than the quality of the ground-state.
However, the high accuracy in the overlaps demonstrates the usefulness of the Born-Oppenheimer concept also for electron-nuclear-photon problems.\\
%%%%%%%%%%%%%%%%%%%%%%%%%%%%%%%%%%%%%%%%%%%%%%%%%%%%%%%%%%%%%%%%%%%%%%%%%%%%%%%%%%%%%%%
% Shin- Metiu subsection
%%%%%%%%%%%%%%%%%%%%%%%%%%%%%%%%%%%%%%%%%%%%%%%%%%%%%%%%%%%%%%%%%%%%%%%%%%%%%%%%%%%%%%%

\section{Cavity QED - The trimer case}
The second system that we analyze is a two-dimensional generalization~\cite{min2014} of the Shin-Metiu model~\cite{shin1996}. The Shin-Metiu model has been analyzed heavily in the context of correlated electron-nuclear dynamics~\cite{albareda2014}, exact forces in non-adiabatic charge transfer~\cite{agostini2015}, or nonadiabatic effects in quantum reactive scattering~\cite{peng2014}, to mention a few. The two-dimensional generalization of the Shin-Metiu model consists of three nuclei and a single electron. Two out of three nuclei are fixed in space. Therefore, this system serves as a model system for a $H_3^+$ molecule that has been confined to two spatial dimensions. In our case, we furthermore place the system into an optical cavity, where it is coupled to a single electromagnetic mode. As Hamiltonian for our system we consider the electron-nuclear part as given in Ref.~\cite{min2014} and couple to the photon Hamiltonian $\hat{H}_p$ from Eq.~[\ref{eq:ham-photons}]. The dipole operator that enters the photon Hamiltonian in Eq.~[\ref{eq:ham-photons}] is given by $\textbf{R} =  \textbf{R}_n -  \textbf{r}_e$, where $\textbf{R}_n$ is the nuclear coordinate and $\textbf{r}_e$ the electronic coordinate. For more details, we refer the reader to (SI5).\\
\begin{figure}[h]
\centerline{\includegraphics[width=0.5\textwidth]{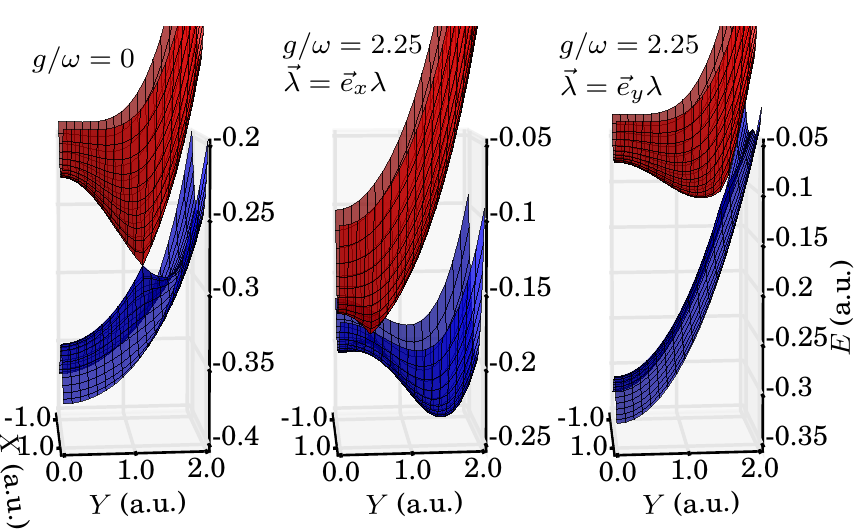}}
\caption{CBO potential energy surfaces for the two-dimensional Shin-Metiu model at $q_\alpha=0$. Increasing matter-photon coupling strength shifts the conical intersection depending on the photon field polarization to larger (smaller) y-values for polarization in y (x)-direction. The plots are using parameters as in Ref.~\cite{min2014}.}
\label{fig:shin-metiu-results}
\end{figure} 
Fig.~\ref{fig:shin-metiu-results} shows the CBO surfaces calculated %using Eq.~[\ref{eq:cavity-boa-surfaces}] 
with $q_\alpha=0$. In all calculations, we tune the matter-photon coupling strength $\lambda$ from the weak-coupling regime to the strong-coupling regime. In the left figure, we choose the value $g/\omega=0$ and in the case of $\lambda = 0$, we find a conical intersection between the first-excited state surface and the second-excited state surfaces as reported in Ref.~\cite{min2014}. In the middle plot, we tune the matter-photon coupling strength to the strong-coupling limit with $g\omega=2.25$ and the photon polarization in x-direction. The matter-photon coupling alters the PES {significantly for} $q_\alpha=0$ and shifts the position of the conical intersection to smaller y-values. The opposite trend can be found if the photon field is polarized in y-direction. As shown in the right plot, for strong coupling with  $g\omega=2.25$, the conical intersection is shifted to larger y-values. These changes of the CBO surfaces have an immediate effect on chemical properties of molecular systems, e.g. the non-adiabatic coupling matrix elements~\cite{kowalewski2016b} that are routinely calculated in nonadiabatic dynamics. These changes in the nonadiabatic coupling terms will affect dramatically the electron-nuclear dynamics and can influence chemical reactions. %$f_{ij}(\textbf{X}) = |\frac{\hbar}{\text{i}} \bra{\phi_i(\textbf{X})}\vec{\nabla}\ket{\phi_j(\textbf{X})}|$ 
To conclude the first three sections, we have seen how the photonic degrees of freedom alter chemical properties of molecular model systems. Besides showing the hybrid character of the ground state and the Rabi splitting of the polaritonic states from first principles, we have identified changes in the BO surfaces that explain, e.g., photon-mediated changes in bondlength or conical intersections.

\section{Local Optimal Control}
\begin{figure}[ht]
\centerline{\includegraphics[width=0.5\textwidth]{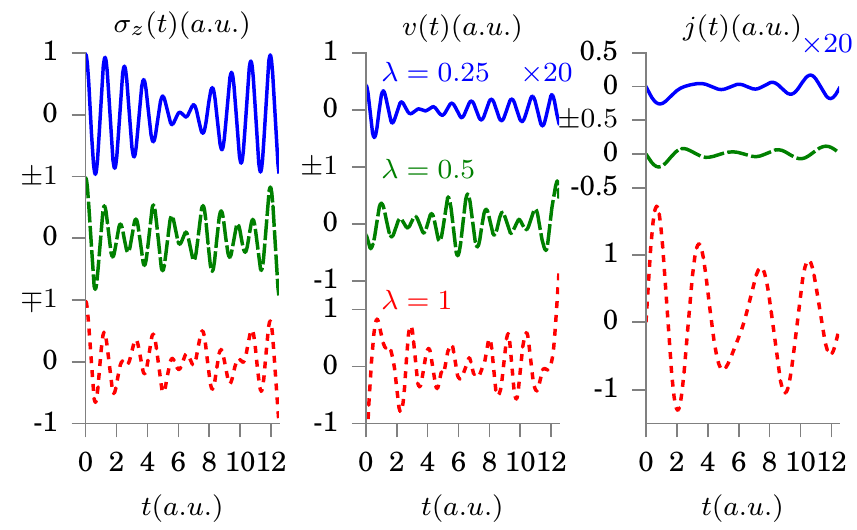}}
\caption{Results for local optimal control: (a) density evolution $\sigma_z(t)$ for $\lambda=0$ in black, $\lambda=0.25$ in blue, $\lambda=0.5$ in green and $\lambda=1.0$ in red and the corresponding evolution of $v(t)$ in (b) and $j(t)$ in (c). Note that $v(t)$ and $j(t)$ for $\lambda=0.25$ has been multiplied by the factor $20$.}
\label{fig:loc-results2}
\end{figure}

While in the first part of this article we have shown how the coupling to photons can alter properties of multi-particle systems, in the second part we investigate which consequences the interaction with photons has in the context of quantum control theory. In quantum control theory we are usually interested in finding an external, classical electromagnetical field, e.g., a specific laser pulse, that forces an electronic system to behave in a previously specified way. Roughly speaking, this can either be done by driving the system into a predefined state with side conditions such as a minimal external field, i.e., optimal control theory~\cite{mancini2005,serban2005,castro2014}, or by prescribing how an observable is supposed to change in time and space, i.e., local control theory~\cite{gross1993,zhu1998,nielsen2013,nielsen2015}. Both approaches can be combined to give local-optimal control theory~\cite{zhu2003, nielsen2015}. We will employ such a hybrid method here. Since quantum control algorithms even for purely electronic systems are numerically very expensive, we will further simplify and consider the simplest yet non-trivial model system of an electron coupled to photons, the extended Rabi model~\cite{shore1993,Braak2011,ruggenthaler2014,pellegrini2015}. %(\Radd{We should name/denote all quantities the same. For instance, we could refer to the previous Hamiltonian for $\hat{q} = const (\hat{a}^{\dagger} + \hat{a})$ and hence connect the parts better.})
\begin{align}
\label{eq:fm-pt-loc}
\hat{H} =& -t_0 \hat{\sigma}_x + \omega \hat{a}^\dagger \hat{a} + \sqrt{\frac{\omega}{2}} \left(\hat{a}^\dagger +\hat{a}\right)\hat{\sigma}_z \nonumber\\
&+ j(t)\left(\hat{a}^\dagger +\hat{a}\right) + v(t)\hat{\sigma}_z,
\end{align}
where $\hat{\sigma}_x$ and $\hat{\sigma}_z$ denote the corresponding Pauli matrices. The Hamiltonian contains as internal parameters the kinetic energy matrix element $t_0$, that yields an amplitude for the electron to hop between the sites, the photon mode frequency $\omega$, that determines the energy of a single photon in the mode, and the electron-photon coupling strength $\lambda$, that fixes the strength of the interaction. Further, Eq.~[\ref{eq:fm-pt-loc}] contains two external variables, which allow us to control the system: the external potential $v(t)$ (corresponding to the usual external laser pulse), which couples to the electron and introduces a potential shift between the sites, and the external dipole $j(t)$, which couples to the photon mode. The external dipole allows to pump the cavity mode. In our calculations, we choose a resonant setup for the three internal parameters, $t_0=2.5$, $\omega=5$ and we vary between $\lambda=0$ (no coupling), {and from the weak- to the strong coupling limit with} $\lambda=\left(0.25,0.5,0.75\right)$.\\
In the following we use the above model to control charge-transfer processes, which are an important topic in the electronic-structure community and have significant implications for, e.g., photovoltaics~\cite{rozzi2013,falke2014}. Similar models have already been used in the same context~\cite{fuks2014}. To see how the coupling to photons changes the charge transfer is motivated also by a recent experiment~\cite{orgiu2015}, where the coupling of an organic semiconductor to photon modes has increased the conductivity by an order of magnitude. To model such a charge-transfer reaction we put most of the charge (why not all will become clear a little later) on one site of our model system and we choose a final time {$T=12.57$} a.u. at which the charge expectation values are interchanged. In terms of the site-basis functions this amounts to  $\ket{\psi(0)} = \ket{0.99,0.01}\, \longrightarrow \, \ket{\psi(T)} = \ket{0.01,0.99}$ where the full initial electron-photon wave function is $\ket{\Psi_0} = \ket{\psi(0)} \otimes \ket{0}$, i.e., the photon mode is initially in the vacuum state. In terms of the charge differences this means we go from $\sigma_z(0) = 0.9{8}$ to $\sigma_z(T) = -0.9{8}$. As a further condition we want to have a minimal external forcing on the electron, which defines the penalty function $P$ by
% \begin{align}
% \label{eq:penalty-functional}
$P = \int_0^T dt \;  v(t)^2$.
% \end{align}
The set up is, however, different to usual optimal control, since we allow to vary the pair $(v(t),j(t))$ to achieve our goal. Clearly, in the case of no coupling $(\lambda =0)$ a change in $j(t)$ will not have any influence on the electronic wave function (since the problem decouples) and so we choose $j(t)=0$. In this case finding the minimum $v(t)$ can be based on an explicit expression of the local control theory~\cite{farzanehpour2012}.
% \begin{align}
% v(t) = -\frac{\frac{1}{2}\ddot{\sigma}_z(t) + 2 t_0^2 \sigma_z(t)}{2\sqrt{t_0^2\left( 1 -\sigma_z^2(t)\right) - \frac{1}{4} \dot{\sigma}^2_z(t) }}.
% \end{align}
The exact expression provides us with a control field for every prescribed path $\sigma_z(t)$ provided the denominator does not go to zero. To avoid such a situation at the initial and final time we have chosen the initial and final states not fully localized. In our approach we use a set of basis functions $\sigma_i(t)=C\cos\left((2i-1)\omega t\right)$ consistent with the initial and final state to expand $\sigma_z(t) = \sum_{i=1}^N c_i \sigma_i(t)$ with $\sum_{i=1}^N c_i = 1$, over which the penalty function $P$ is minimized, i.e., $\min_{\{c_i \}}\int_0^T dt \; v([\sigma_z];t)^2$. In all calculations we use {N=11}. Note that such an expansion is simple only in the case of physical observables such as the charge difference $\sigma_z(t)$, while in terms of time-dependent wave functions $\ket{\Psi(t)}$ this is extremely demanding. The possibility to restrict to a simple and finite basis of charge paths $\sigma_i(t)$ is one advantage of this local optimal control approach. The other advantage is that the charge transfer is guaranteed to be achieved at the final time. The minimization is then performed by the quasi-Newton method~\cite{broyden1970}. Extending the number of basis functions would lead to an even further optimized value of the penalty function $P$, but does not lead to qualitative differences in the discussion of the obtained results.\\ %, e.g., the simplex method~\cite{nelder1965} and
In the case of $\lambda \neq 0$ we do not have a simple analytical expression for $v(t)$ and the electronic part of the wave function will depend also on the choice of external dipole $j(t)$. To find the corresponding $v([\sigma_z,j];t)$ for a given density path $\sigma_z(t) = \sum_{i=1}^N c_i \sigma_i(t)$ and dipole $j(t) = \sum_{k=1}^M d_k j_k(t)$, where we choose $j_k(t) = \sin\left(k \omega t\right)$~\footnote{We point out, that due to the simple connection between $j(t)$ and $q(t)=\langle \hat{a}^{\dagger} + \hat{a} \rangle$ via the (one-mode) Maxwell's equation $\partial_t^2 q(t) + \omega^2 q(t)=-2 \omega (\lambda \sigma_z(t) + j(t))$, we can directly consider an expansion in terms of the external dipole instead of $q(t)$. This is convenient since we do not care in this example about the mode occupation but about keeping $j(t)$ relatively small and choose an expansion accordingly.}, we use a fixed-point method~\cite{ruggenthaler2011b,nielsen2013,ruggenthaler2014,flick2015}. Now we can vary over a space of electronic $c_i$ and photonic $d_k$ coordinates, i.e.
%\begin{align}
$\min\limits_{\{c_i, d_k \}}\int_0^T dt \; v([\sigma_z, j];t)^2$,
%\end{align}
where we choose {$M=11$}. We first briefly review the trivial cases, where no external potentials, i.e. $v(t) = j(t) = 0$, are applied to the system. Tab.~\ref{tab:loc:psi_final} shows the values for $\sigma_z(T)$ for all four cases of the electron-photon coupling strength $\lambda$. While $\sigma_z(0)$ is given by the initial state and thus equals $0.980$ in all examples, they vary strongly in their free final state $\sigma_z(T)$. Analyzing these values already gives us the first hint, how to optimize electron-photon problems to our favor. If there is no electron-photon coupling ($\lambda=0$), the final value for $v(t)=0$ is $\sigma_z(T)=0.98$. In contrast, if we choose the coupling strength of $\lambda=0.25$, the free evolution without any external potential already yields a final state very close to the desired value $-0.98$. Thus, we can conclude from this observation that already tuning the electron-photon coupling strength $\lambda$ allows to utilize the electron-photon coupling. Additional, we can suggest that a local control optimization for $\lambda=0.25$ coupling strength can be very efficient, since the external potential has to improve the outcome only little. In contrast, for all other values of $\lambda$ we see that the external potential has to modify the evolution more strongly.
\begin{table}
\begin{tabular}{ c| r | r |r | r} 
 $g=\sqrt{\frac{\omega}{2}}\lambda$ &  $0$ & $0.25$ & $0.5$ & $1.0$\\\hline
%$\sigma_z(0)$& $0.980$& $0.980$& $0.980$& $0.980$\\
$v(t) =  j(t) = 0:$ $\sigma_z(T)$& $0.980$ & $-0.979$ & $0.975$& $0.969$\\\hline
%$v(t) \neq 0:$ $\sigma_z(T)$& $-0.980$ & $-0.980$ & $-0.980$& $-0.980$\\\hline
$v(t) \neq 0, j(t) = 0:$ $P$& $0.3015$ & $0.0008$ & $0.8145$ & $ 3.2473$ \\
$v(t) \neq 0, j(t) \neq 0:$ $P$& $-$ & $0.0008$ & $0.8113$ & $2.4084$
\end{tabular}
\caption{Results of the local control optimization for the extended Rabi model.}
\label{tab:loc:psi_final}
\end{table}
\begin{figure}[h]
\centerline{\includegraphics[width=0.5\textwidth]{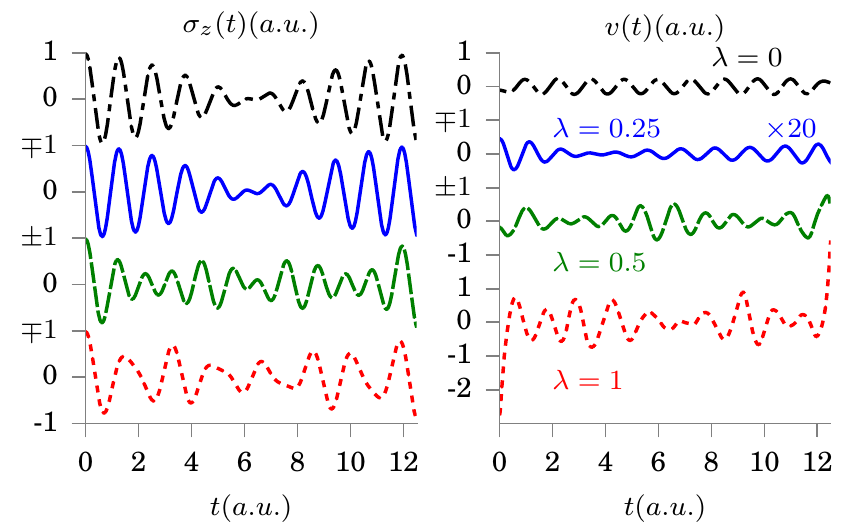}}
\caption{Results for local optimal control: (a) density evolution $\sigma_z(t)$ for $\lambda=0$ in black, $\lambda=0.25$ in blue, $\lambda=0.5$ in green and $\lambda=1.0$ in red and the corresponding evolution of $v(t)$. Note that $v(t)$ for $\lambda=0.25$ has been multiplied by the factor $20$.}
\label{fig:loc-results1}
\end{figure}
The resulting external potentials of the local control optimizations with fixed $j(t) = 0$ are shown in Fig.~\ref{fig:loc-results1}. For the case of $\lambda=0$, we find a rather regular oscillation in $v(t)$ and $\sigma_z(t)$. The evolution of $\sigma_z(t)$ for the case of $\lambda=0.25$ is very close to the optimal evolution of $\lambda=0$, but has a very small value of $v(t)$, due to the optimal utilization of the electron-photon interaction. For the cases of $\lambda=0.5$ and $\lambda=1.0$ the electron-photon interaction is stronger, thus the system reacts stiffer with respect to the external potential $v(t)$. This leads to a higher penalty function as shown in Tab.~\ref{tab:loc:psi_final}. For these two examples, we further find a non-symmetric optimal solution of $v(t)$. This can be explained by the fact, that requesting a final value of $\sigma_z(T) = -0.98$ does not give restrictions on the final photon state. Here, we find excitations of the photonic amplitude.
We now turn our focus to the optimization where we lift the restriction on $j(t)$. In Fig.~\ref{fig:loc-results2}, and the last row in Tab.~\ref{tab:loc:psi_final}, we show the optimization for the case of $j(t)\neq0$. Here we find that the additional degree of freedom allows us to control the system more efficiently. In particular the example of $\lambda=1$ shows the effectiveness of the scheme. Here, we are able to lower the penalty function significantly from $3.2473$ for $j(t)=0$ to $2.4084$ for $j(t)\neq0$.
Our local optimal control results for this simple model show how the coupling to photons can induce charge transfer reactions with only little external forcing on the electron. While it is not surprising that in a resonant set up the reaction is driven mainly by the coupling to the mode, the optimal-control analysis shows that controlling the photonic part of the electron-photon wave function via $j(t)$ directly allows a reduction of the applied external potential $v(t)$ that acts on the electron. This indicates the possiblity to optimize charge-transfer reactions in a cavity by specifically populating certain cavity modes via an external dipole or current. This is different to the usual control approaches where one controls the electronic systems via an external laser only.

\section{Quantum-electrodynamical density-functional theory}

The last section of this paper is dedicated to quantum-electrodynamical density-functional theory (QEDFT). It allows for numerically feasible \textit{ab-initio} simulations of correlated matter-photon systems. The basic idea is that instead of solving for the (usually infeasible) correlated electron-photon wave function one solves a set of self-consistent (in practice approximate) equations of motion for specific reduced quantities. For details on the method we refer to~\cite{ruggenthaler2011b, tokatly2013, ruggenthaler2014, pellegrini2015, flick2015}. {All available implementations of QEDFT~\cite{ruggenthaler2014, pellegrini2015, flick2015} are based on the electron density as basic variable. However, a consistent treatment of the quantized electric and magnetic field beyond the dipole coupling is possible by using QED-Current-DFT~\cite{ruggenthaler2014}.} In this article, we consider the performance of QEDFT for an approximation based on the OEP scheme~\cite{pellegrini2015} for the case of single-photon emission and bound electron-photon states. We compare the semi-classical (mean-field) and the OEP approximation to the exact numerical treatment beyond the rotating-wave approximation (RWA)~\cite{shore1993} for a simple model.\\
In contrast to the previous models we now consider many photon modes that couple to our particle system. To that end we apply the model Hamiltonian introduced in Ref.~\cite{buzek1999} but go beyond the RWA. We consider an electronic two-level systems coupled to $M=400$ modes. In order to be able to treat the photon field consisting of $M$ modes numerically exactly we will truncate the Fock space and only consider the vacuum state, the $M$ one-photon states and the $\left(M^2-M\right)/2$ two-photon states in a (1D) cavity of volume (length) $V$. The Hamiltonian we employ is given by~\cite{buzek1999}
\begin{align}
  \label{eq:buzek-ham}
  \hat{H} & = -t_0 \hat{\sigma}_x + \sum_{\alpha} \omega_{\alpha} \hat{a}_{\alpha}^{\dagger} \hat{a}_{\alpha} + \sum_{\alpha} \omega_\alpha \lambda_\alpha \hat{q}_{\alpha}  \left(d_{eg}\hat{\sigma}_z\right) 
\end{align}
where $\hat{q}_{\alpha}$ as in Eq.~[\ref{eq:ham-photons}] and the wave vectors $k_\alpha = \omega_\alpha/c = \alpha \pi/V$. We fix the position of the two-level subsystems at $x=V/2$ and hence we can deduce the coupling constants from the photon modes $\lambda_\alpha(x) = \sqrt{\frac{2}{\hbar\,\epsilon_0\,V} } \, \sin(k_{\alpha} x)$
at this position. The quantized {electric field} is then given by $\hat{E}(x)=\sum_{\alpha}\omega_\alpha\lambda_\alpha(x)\hat{q}_\alpha$, while the quantity that is linked more closely to the quantum nature of the light field is the intensity observable~\cite{buzek1999} that is given by $\langle\hat{E}^2(x,t)\rangle= \sum_{\alpha,\beta}{\omega_\alpha\omega_\beta}\lambda_\alpha(x)\lambda_\beta(x)\left<\hat{q}_\alpha(t)\hat{q}_\beta(t)\right>.$ As parameters for the two-level system, we use a one-dimensional Hydrogen atom with a soft-Coulomb potential. We consider the first two levels of such a system and employ the parameters as in Ref.~\cite{su1991}. Thus, $t_0=0.197$, $d_{eg}=1.034$, and {$\lambda_\alpha=\pm 0.0103$}. In the following, we discuss two different initial states. The setup (\textbf{1}) features the intial state  $\ket{\Psi(t_0)} = \ket{e}\otimes \ket{0}$, where $\ket{e}$ is the excited state of the bare electronic Hamiltonian of Eq.~[\ref{eq:buzek-ham}] and $\ket{0}$ indicates the photon field in the vacuum state. During time-evolution the electronic excitation will decay to the ground state and hereby emit a single photon via spontaneous emission~\cite{buzek1999}. This corresponds to the classical textbook case, except that we treat our system beyond the RWA. In the second case (\textbf{2}), the setup to analyze the single-photon emission process consists of an factorizable initial state $\ket{\Psi(t_0)} = \left(\sqrt{\frac{1}{500}}\ket{s_1} + \sqrt{\frac{499}{500}}\ket{s_2}\right)\otimes \ket{0},$ where $\ket{s_1}$ and $\ket{s_2}$ refer to the individual sites of the two-site model.
\begin{figure}[h]
\includegraphics[width=0.5\textwidth]{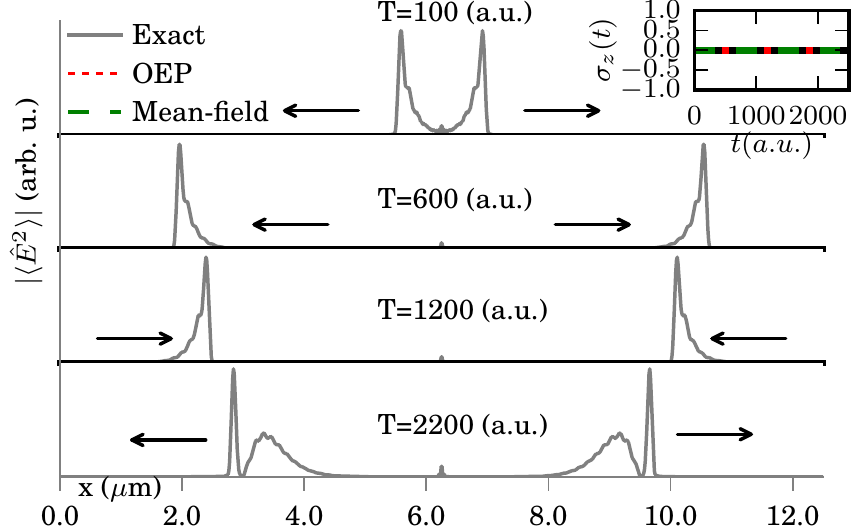}
\caption{Time-evolution of the spontaneous-photon emission process: Expectation value of the absolute of the intensity-field operator $\braket{\hat{E}^2(t)}$ in the exact simulation (gray), the OEP approximation (red) and the semi-classical approximation (green) at time T=100 a.u., 600 a.u., 1200 a.u. and 2200 a.u.. The inset shows the dipole moment $\braket{\sigma_z(t)}$.}\label{fig:spon-em-efield-psi0}
\end{figure}
We start by discussing the dipole moment of the system, i.e., $\left<\sigma_z(t)\right>$. The QEDFT reformulation of Eq.~[\ref{eq:buzek-ham}] has the basic functional variables $(\left<\hat{\sigma}_z(t)\right>,\{\left<\hat{q}_\alpha(t)\right>\})$~\cite{ruggenthaler2014}, which makes this quantity specifically simple to determine. In the inset of Fig.~\ref{fig:spon-em-efield-psi0}, we show the time-evolution of $\left<\sigma_z(t)\right>$. We find that in the exact propagation the dipole moment $\sigma_z(t)=0$, i.e., the deexcitation from the excited state to the ground state of the atom is a dipole-free transition. This implies that also the electric field observable in this process is zero for all times ($E(x,t)=0$). However, as shown in Fig.~\ref{fig:spon-em-efield-psi0}, the intensity of the spontaneous emission for this process is nonzero~\cite{buzek1999}. At initial time, we find two sharp wave fronts appearing, which travel to the boundaries, are reflected at the cavity mirrors and excite the atom again. {The semi-classical approximation and the OEP approximation for this setup correctly reproduce the (trivial) dipole-moment and electric field, which are the basic variables. However, simple approximations to the intensity evolution that use the $\sigma_z(t)$ and $d_{12}(t)$ of the QEDFT systems fail in correctly describing the intensity evolution. {Since $\sigma_z(t)$ and $d_{12}(t)$ are equal to zero for all times, the exact functional for the intensity has to provide the correct time evolution of the observable exclusively through the dependence on the initial state.} This is one of the drawbacks of an implicit functional reformulation of quantum physics, where we do not know the explicit forms of all observables but are often dependent on simple approximations. That these approximations can be useful, though, will become clear in the next case.} For this example, 
\begin{figure}[h]
\includegraphics[width=0.5\textwidth]{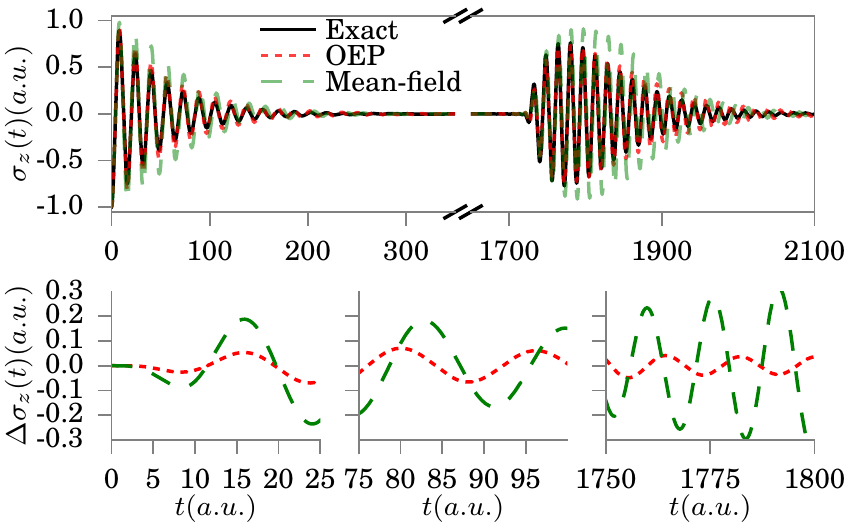}
\caption{Time-evolution of the single-photon emission process: Upper panel: expectation value of the dipole-moment operator $\braket{\hat{\sigma}_z(t)}$ in the exact simulation (black), the OEP approximation (red) and the semi-classical mean-field approximation (green). The lower panel shows the difference $\Delta \sigma_z(t)$ of  OEP approximation (red) and the semi-classical mean-field approximation (green) to the exact propagation at different insets of the full evolution.}
\label{fig:spon-em-sigz}
\end{figure} 
the time-dependent evolution of the dipole moment for the two-site model is shown in the upper panel of Fig.~\ref{fig:spon-em-sigz}. Here, we find an initial exponential decay of the dipole oscillation that is the single-photon emission of the atom. After t=1800 a.u., we find the re-absorption of the emitted photon and the dipole moment starts to oscillate again. The exact simulation is shown in black. Our approximate QEDFT propagation based on the OEP approximation, shown in red, is very close to the exact results as can be seen in the second row of the figure. The mean-field approximation also performs qualitatively correct. It is capable of reproducing the emission process and also the re-absorption of the photon. However it misses some quantitative features. The emission time is too long, which means that the photon in the exact simulation is emitted faster. The same can be seen for the re-absorption of the photon. Here, the mean-field dipole moment evolution is broader than the exact and the OEP approximation.
\begin{figure}[h]
\includegraphics[width=0.5\textwidth]{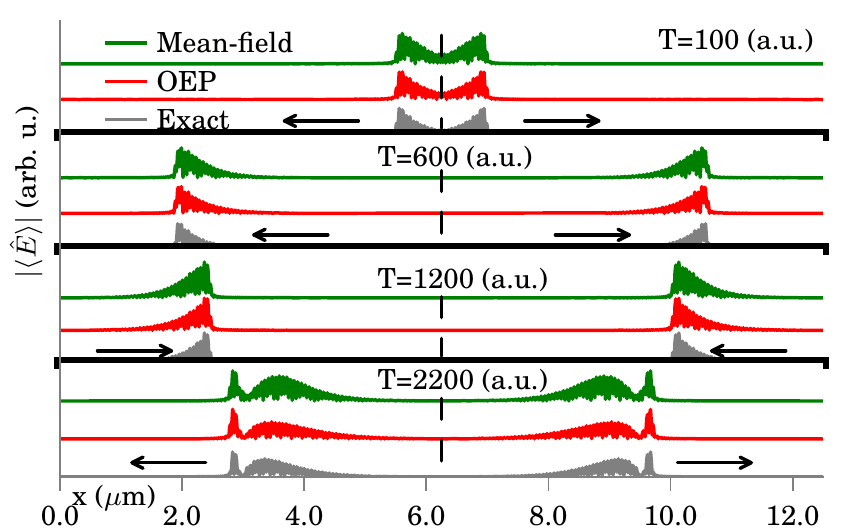}
\caption{Time-evolution of the single-photon emission process: Expectation value of the absolute of the {electric-field} operator $\braket{\hat{E}(t)}$ in the exact simulation (gray), the OEP approximation (red) and the semi-classical approximation (green) at time T=100 a.u., 600 a.u., 1200 a.u. and 2200 a.u.}\label{fig:spon-em-efield}
\end{figure} 
{In Fig.~\ref{fig:spon-em-efield}, we} plot the absolute value of the expectation value of the {electric field} operator. We observe after T=100 a.u. a wave packet with a sharp front travelling towards the boundaries of the cavity. After T=1200 a.u. the wave packets are reflected by the boundary and they travel back to the atom, where they are re-absorbed and then re-emitted into the field again. This process generates a second maxima in the wave packet that can be observed in the third column of Fig.~\ref{fig:spon-em-efield}. The shapes of the wave packet in the OEP approximation, shown in red nicely agree with the exact shapes, here shown in grey. The mean-field approximation is again qualitatively accurate, but in particular the second maxima is too broad due to the wrong decay time of the two-level system.
\begin{figure}[h]
\includegraphics[width=0.5\textwidth]{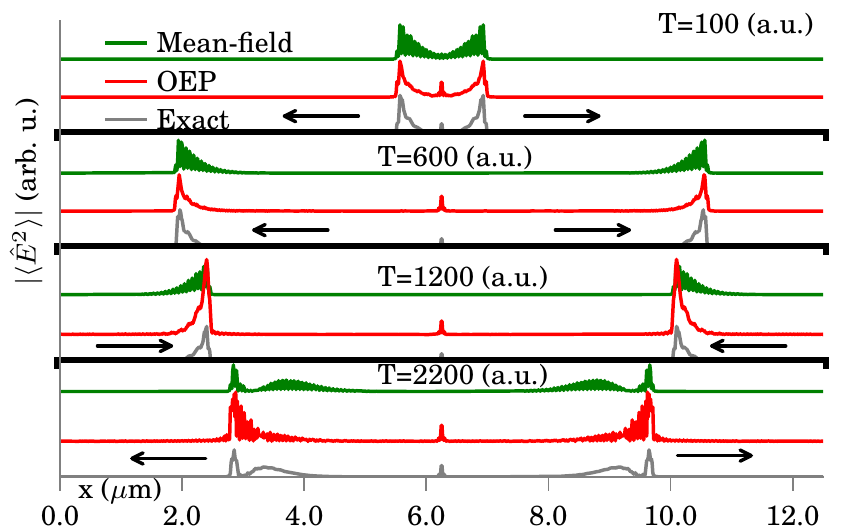}
\caption{Time-evolution of the single-photon emission process: Expectation value of the absolute of the intensity-field operator $\braket{\hat{E}^2(t)}$ in the exact simulation (gray), the OEP approximation (red) and the semi-classical approximation (green) at time T=100 a.u., 600 a.u., 1200 a.u. and 2200 a.u.}\label{fig:spon-em2-efield}
\end{figure}
In Fig.~\ref{fig:spon-em2-efield}, we plot the absolute value of $\braket{\hat{E}^2(t)}$ {with the same color-coding as before. Here, while the OEP develops unphysical oscillations in the wavefront after reflection at the mirrors, it is able to reproduce a remaining photon intensity at the position of the atom.} This effect is a beyond semi-classical and a two-photon effect, which is also missed by the RWA~\cite{buzek1999}. Since in the semi-classical approximation the matter and the photon system completely decouple, no intensity remains at the position of the atom. In the exact and also the OEP approximation however the systems are still correlated leading to the remaining intensity. Indeed, this intensity is due to the hybrid ground state of the correlated electron-photon system and hence corresponds to a bound electron-photon state. This comparison allows us to conclude that we successfully identified important beyond semi-classical effects, that can be described by a QEDFT approximation.

\section{Summary and Conclusion}

In summary, in this paper we have illustrated how long-standing concepts of quantum chemistry have to be adapted if the electron-photon interaction is considered in the quantum limit. We have reported the adapted concept of Cavity-Born Oppenheimer surfaces, that we calculated for a dimer system and the Shin-Metiu model. This concept is accurate for static calculations from the weak- to the strong coupling limit and can be used to predict chemical quantities such as bond length, non-adiabatic coupling terms or absorption spectra. In all examples, we compared the approximate solutions to the numerical exact solutions. In a local control scheme, we have shown how we can use the electron-photon interaction to our favor to modify chemical reactions more efficiently. The additional degree of freedom in the photon subsystem offers new promising possibilities. In the last section, we have shown how a density-functional approach can be superior to the semi-classical approach for bound polariton states. These states appear in optical cavities and require a correct description of the correlated electron-photon interaction. This work on the interface of quantum optics and material science impacts both research fields and can lead to novel applications in chemistry and material science, such as new photonic devices or laser technologies.\\

We acknowledges financial support from the European Research Council(ERC-2015-AdG-694097), Spanish grant (FIS2013-46159-C3-1-P), Grupos Consolidados (IT578-13), and AFOSR Grant No. FA2386-15-1-0006 AOARD 144088, H2020-NMP-2014 project MOSTOPHOS (GA no. 646259)
and COST Action MP1306 (EUSpec) and the Austrian Science Foundation (FWF P25739-N27).

%%%%%%%%%%%%%%%%%%%%%%%%%%%%%%%%%%%%%%%%%%%%%%%%%%%%%%%%%%%%%%%%%%
%                        Bibliography                            %
%%%%%%%%%%%%%%%%%%%%%%%%%%%%%%%%%%%%%%%%%%%%%%%%%%%%%%%%%%%%%%%%%%
\bibliographystyle{apsrev4-1}
\bibliography{qed_chemistry} % Produces the bibliography via BibTeX.

%%%%%%%%%% Merge with supplemental materials %%%%%%%%%%
\pagebreak
\widetext
\begin{center}
\textbf{\large Supplemental Information:\\Atoms and Molecules in Cavities: From Weak to Strong Coupling in QED Chemistry}
\end{center}
%%%%%%%%%% Merge with supplemental materials %%%%%%%%%%
%%%%%%%%%% Prefix a "S" to all equations, figures, tables and reset the counter %%%%%%%%%%
\setcounter{equation}{0}
\setcounter{figure}{0}
\setcounter{table}{0}
\setcounter{page}{1}
\makeatletter
\renewcommand{\theequation}{S\arabic{equation}}
\renewcommand{\thefigure}{S\arabic{figure}}
%\renewcommand{\bibnumfmt}[1]{[S#1]}
%\renewcommand{\citenumfont}[1]{S#1}
%%%%%%%%%% Prefix a "S" to all equations, figures, tables and reset the counter %%%%%%%%%%

\subsection{(SI1) Relative Jacobi coordinates for four-body systems:}
The Hamiltonian [1-4] can be written explicitely for the studied dimer system as follow
\begin{align}
\label{eq:qed-chemistry-hamiltonian-appendix1}
\hat{H} &= \hat{H}_{en} + \hat{H}_p\\ 
\hat{H}_{en} &= -\frac{\hbar^2}{2M_1}\vec{\nabla}_{\textbf{X}_1}^2 -\frac{\hbar^2}{2M_2}\vec{\nabla}_{\textbf{X}_2}^2  -\frac{\hbar^2}{2m_3}\vec{\nabla}_{\textbf{x}_3}^2   -\frac{\hbar^2}{2m_4}\vec{\nabla}_{\textbf{x}_4}^2 \nonumber\\
&+\frac{Z_1Z_2e^2}{4\pi\epsilon_0\sqrt{\left(\textbf{X}_1-\textbf{X}_2\right)^2+1}} -  \frac{Z_1e^2}{4\pi\epsilon_0\sqrt{\left(\textbf{X}_1-\textbf{x}_3\right)^2+1}}\nonumber\\
&- \frac{Z_1e^2}{4\pi\epsilon_0\sqrt{\left(\textbf{X}_1-\textbf{x}_4\right)^2+1}}-\frac{Z_2e^2}{4\pi\epsilon_0\sqrt{\left(\textbf{X}_2-\textbf{x}_3\right)^2+1}} \nonumber\\
&-  \frac{Z_2e^2}{4\pi\epsilon_0\sqrt{\left(\textbf{X}_2-\textbf{x}_4\right)^2+1}} + \frac{e^2}{4\pi\epsilon_0\sqrt{\left(\textbf{x}_3-\textbf{x}_4\right)^2+1}}\\
\hat{H}_p &= \frac{1}{2}\sum\limits_\alpha\left[\hat{p}^2_\alpha+\omega_\alpha^2\left(\hat{q}_\alpha + \frac{\boldsymbol \lambda_\alpha} {\omega_\alpha} \cdot{e\textbf{R}} \right)^2\right]\label{eq:ham-photons-appendix1}\\
{\textbf{R}} & = Z_1\textbf{X}_1 + Z_2\textbf{X}_2 - \textbf{x}_3 - \textbf{x}_4
\end{align} 
Here, the capital variables, $\textbf{X}_1$ and $\textbf{X}_2$, denote the nuclear coordinates, while the small variables $\textbf{x}_3$ and $\textbf{x}_4$ denote the electronic coordinates. In the following, we briefly want to discuss the relative coordinates and the real-space grid used for the numerical calculations of section 1. The coordinates are the electron distance coordinate $\textbf{x} = \textbf{x}_1-\textbf{x}_2$, the nuclear distance coordinate $\textbf{X} = \textbf{X}_1-\textbf{X}_2$, the distance between the electronic and the nuclear center of masses ${\boldsymbol\xi}$ and the global center of mass $\textbf{X}_{CM2}$~\cite{crawford2015}. In this new coordinate system the Hamiltonians of Eqs.~[\ref{eq:qed-chemistry-hamiltonian-appendix1}]-[\ref{eq:ham-photons-appendix1}] and the dipole moment become
\begin{align}
\label{eq:qed-chemistry-hamiltonian-appendix2} 
\hat{H}_{en} &= -\hbar^2\frac{M_1+M_2}{2M_1M_2}\vec{\nabla}_{\textbf{X}}^2 -\hbar^2\vec{\nabla}_{\textbf{x}}^2  -\hbar^2\frac{2+M_1+M_2}{4(M_1+M_2)}\vec{\nabla}_{{\boldsymbol\xi}}^2  -\frac{\hbar^2}{2\left(2+M_1+M_2\right)}\vec{\nabla}_{\textbf{x}_\text{CM2}}^2 \nonumber\\
&+\frac{Z_1Z_2e^2}{4\pi\epsilon_0\sqrt{\textbf{X}^2+1}}+ \frac{e^2}{4\pi\epsilon_0}{\sqrt{\textbf{x}^2+1}}\nonumber\\ 
&- \frac{Z_1e^2}{4\pi\epsilon_0\sqrt{\left({\boldsymbol \xi}+\frac{\textbf{x}}{2}+\frac{M_2\textbf{X}}{M_1+M_2}\right)^2+1}} -\frac{Z_2e^2}{4\pi\epsilon_0\sqrt{\left(-{\boldsymbol \xi}+\frac{\textbf{x}}{2}+\frac{M_1\textbf{X}}{M_1+M_2}\right)^2+1}} \nonumber\\
&- \frac{Z_2e^2}{4\pi\epsilon_0\sqrt{\left(-{\boldsymbol \xi}-\frac{\textbf{x}}{2}+\frac{M_1\textbf{X}}{M_1+M_2}\right)^2+1}}- \frac{Z_1e^2}{4\pi\epsilon_0\sqrt{\left({\boldsymbol \xi}-\frac{\textbf{x}}{2}+\frac{M_2\textbf{X}}{M_1+M_2}\right)^2+1}}\\
\hat{H}_p &= \frac{1}{2}\sum\limits_\alpha\left[\hat{p}^2_\alpha+\omega_\alpha^2\left(\hat{q}_\alpha + \frac{\boldsymbol \lambda_\alpha} {\omega_\alpha} \cdot{\textbf{R}} \right)^2\right]\\
{\textbf{R}} & =-2{\boldsymbol\xi}+\frac{\textbf{X}\left(M_1Z_2-M_2Z_1\right)}{M_1+M_2}
\end{align}  

\subsection{(SI2) General Cavity Born-Oppenheimer for correlated electron-photon systems}
The CBO approximation to the Hamiltonian [1-4] in the paper can be constructed as follows: First solve the electronic Schr\"odinger equation
\begin{align}
&\hat{H}_e(\left\{\textbf{X}\right\},\{q_\alpha\}) \; {\phi_j(\left\{\textbf{X}\right\},\{q_\alpha\})}  =E_j(\left\{\textbf{X}\right\},\{q_\alpha\}) \; {\phi_j(\left\{\textbf{X}\right\},\{q_\alpha\}})
\end{align}
with the electronic Hamiltonian
\begin{align}
\label{eq:ham-cboa-full}
&\hat{H}_e(\left\{\textbf{X}\right\},\{q_\alpha\})= \hat{T}_e + \hat{W}_{ee} + \hat{W}_{en}(\left\{\textbf{X}\right\}) + \sum\limits_\alpha\frac{1}{2}\left({\boldsymbol \lambda_\alpha}\cdot \textbf{R} \right)^2+\sum\limits_\alpha \omega_\alpha{q}_\alpha {\boldsymbol \lambda_\alpha}  \cdot{\textbf{R}} 
\end{align}
In a second step we then solve the nuclear-photon problem
\begin{align}
\label{eq:qed-chemistry-nuclearCBOAHamiltonian-full}
\hat{H}_j(\left\{\textbf{X}\right\},\{q_\alpha\}) =  \hat{T}_n  +\sum\limits_\alpha\frac{1}{2}\hat{p}^2_\alpha + V_j(\left\{\textbf{X}\right\},\{q_\alpha\})
\end{align}
with the potential-energy surfaces (PES) determined from the electronic part that are given by
\begin{align}
V_j(\left\{\textbf{X}\right\},\{q_\alpha\}) &= E_j(\left\{\textbf{X}\right\},\{q_\alpha\})+\hat{W}_{nn}(\left\{\textbf{X}\right\}) +\sum\limits_\alpha\frac{1}{2} \omega_\alpha^2{q}^2_\alpha.
\label{eq:cavity-boa-surfaces-full}
\end{align}
This procedure allows us to effectively decouple the system into an electronic part and a nuclear-photon part. Using 
\begin{align}
&\hat{H}_j(\left\{\textbf{X}\right\},\{q_\alpha\}){\chi_{ij}(\left\{\textbf{X}\right\},\{q_\alpha\})} = \epsilon_i(\left\{\textbf{X}\right\},\{q_\alpha\}) {\chi_{ij}(\left\{\textbf{X}\right\},\{q_\alpha\}}),
\end{align}
where ${\chi_{ij}}$ denotes a correlated nuclear-photon wave function. Having all necessary wave functions at hand, we can construct the BO states, i.e. for the BO ground state $\ket{\Psi_0} = \ket{\chi_{00}}\ket{\phi_0}$.

\subsection{(SI3) Cavity Born-Oppenheimer for four-body systems:}
To solve the BO problem for the dimer system, we first compute the electronic part that for this system parametrically depends on the position of the nuclei $\textbf{X}_1$, and $\textbf{X}_2$, and on the displacement coordinates of the photon modes $\{q_\alpha\}$, i.e. we solve the following electronic equation
\begin{align}
&\hat{H}_e(\textbf{X}_1,\textbf{X}_2,\{q_\alpha\}) \; {\phi_j(\textbf{X}_1,\textbf{X}_2,\{q_\alpha\})}  =E_j(\textbf{X}_1,\textbf{X}_2,\{q_\alpha\}) \; {\phi_j(\textbf{X}_1,\textbf{X}_2,\{q_\alpha\}})
\end{align}
with the electronic Hamiltonian
\begin{align}
\label{eq:ham-cboa}
&\hat{H}_e(\textbf{X}_1,\textbf{X}_2,\{q_\alpha\}) \nonumber\\
&= -\frac{\hbar^2}{2m_3}\vec{\nabla}_{\textbf{x}_3}^2   -\frac{\hbar^2}{2m_4}\vec{\nabla}_{\textbf{x}_4}^2 \nonumber\\
& + \frac{e^2}{4\pi\epsilon_0\sqrt{\left(\textbf{x}_3-\textbf{x}_4\right)^2+1}}-  \frac{Z_1e^2}{4\pi\epsilon_0\sqrt{\left(\textbf{X}_1-\textbf{x}_3\right)^2+1}}\nonumber\\
&- \frac{Z_1e^2}{4\pi\epsilon_0\sqrt{\left(\textbf{X}_1-\textbf{x}_4\right)^2+1}}-\frac{Z_2e^2}{4\pi\epsilon_0\sqrt{\left(\textbf{X}_2-\textbf{x}_3\right)^2+1}} \nonumber\\
&-  \frac{Z_2e^2}{4\pi\epsilon_0\sqrt{\left(\textbf{X}_2-\textbf{x}_4\right)^2+1}} + \sum\limits_\alpha\frac{1}{2}\left({\boldsymbol \lambda_\alpha}\cdot \textbf{R} \right)^2+\sum\limits_\alpha \omega_\alpha{q}_\alpha {\boldsymbol \lambda_\alpha}  \cdot{\textbf{R}} 
\end{align}
In this way, the electronic wavefunctions ${\phi_j(\textbf{X}_1,\textbf{X}_2,\{q_\alpha\}})$ have parametrical dependency on the nuclear and photonic coordinates. In a second step we then solve the nuclear-photon problem
\begin{align}
\label{eq:qed-chemistry-nuclearCBOAHamiltonian}
\hat{H}_j(\textbf{X}_1,\textbf{X}_2,\{q_\alpha\}) =  -\frac{\hbar^2}{2M_1}\vec{\nabla}_{\textbf{X}_1}^2 -\frac{\hbar^2}{2M_2}\vec{\nabla}_{\textbf{X}_2}^2  
 +\sum\limits_\alpha\frac{1}{2}\hat{p}^2_\alpha + V_j(\textbf{X}_1,\textbf{X}_2,\{q_\alpha \})
\end{align}
with the potential-energy surfaces determined from the electronic part that are given by
\begin{align}
V_j(\textbf{X}_1,\textbf{X}_2,\{q_\alpha \}) &= E_j(\textbf{X}_1,\textbf{X}_2,\{q_\alpha \})+\frac{Z_1Z_2e^2}{4\pi\epsilon_0\sqrt{\left(\textbf{X}_1-\textbf{X}_2\right)^2+1}} +\sum\limits_\alpha\frac{1}{2} \omega_\alpha^2{q}^2_\alpha.
\label{eq:cavity-boa-surfaces}
\end{align}
This procedure allows us to effectively decouple the system into an electronic part and a nuclear-photon part. Using 
\begin{align}
&\hat{H}_j(\textbf{X}_1,\textbf{X}_2,\{q_\alpha \}){\chi_{ij}(\textbf{X}_1,\textbf{X}_2,\{q_\alpha \})} = \epsilon_i(\textbf{X}_1,\textbf{X}_2,\{q_\alpha \}) {\chi_{ij}(\textbf{X}_1,\textbf{X}_2,\{q_\alpha \}}),
\end{align}
where ${\chi_{ij}}$ denotes a correlated nuclear-photon wave function, we can construct the BO states, i.e. for the BO ground state $\ket{\Psi_0} = \ket{\chi_{00}}\ket{\phi_0}$.
%Next we determine the exact results based on the Hamiltonian of Eq.~[\ref{eq:qed-chemistry-hamiltonian}] and compare it to the CBOA for different values of ${{\boldsymbol \lambda_\alpha}}$. 
The electronic Hamiltonian that parametrically depends on $\textbf{X}$ and $\{q_\alpha\}$ of Eq.~[\ref{eq:ham-cboa}] 
becomes
\begin{align}
\label{eq:qed-chemistry-hamiltonian-appendix2}
\hat{H}_{e}(\textbf{X},\{q_\alpha\}) &= -\hbar^2\vec{\nabla}_{\textbf{x}}^2  -\frac{\hbar^2}{4}\vec{\nabla}_{{\boldsymbol\xi}}^2  \nonumber\\ &+\frac{Z_1Z_2e^2}{4\pi\epsilon_0\sqrt{\textbf{X}^2+1}}+ \frac{e^2}{4\pi\epsilon_0}{\sqrt{\textbf{x}^2+1}}\\ 
&- \frac{Z_1e^2}{4\pi\epsilon_0\sqrt{\left({\boldsymbol \xi}+\frac{\textbf{x}}{2}+\frac{M_2\textbf{X}}{M_1+M_2}\right)^2+1}} -\frac{Z_2e^2}{4\pi\epsilon_0\sqrt{\left(-{\boldsymbol \xi}+\frac{\textbf{x}}{2}+\frac{M_1\textbf{X}}{M_1+M_2}\right)^2+1}} \nonumber\\
&- \frac{Z_2e^2}{4\pi\epsilon_0\sqrt{\left(-{\boldsymbol \xi}-\frac{\textbf{x}}{2}+\frac{M_1\textbf{X}}{M_1+M_2}\right)^2+1}} - \frac{Z_1e^2}{4\pi\epsilon_0\sqrt{\left({\boldsymbol \xi}-\frac{\textbf{x}}{2}+\frac{M_2\textbf{X}}{M_1+M_2}\right)^2+1}}\\
&+ \sum\limits_\alpha\frac{1}{2}\left({\boldsymbol \lambda_\alpha}\cdot \textbf{R} \right)^2 +\sum\limits_\alpha \omega_\alpha{q}_\alpha {\boldsymbol \lambda_\alpha}  \cdot{\textbf{R}} \nonumber
\end{align}
The nuclear-photon problem of the CBO approximation Eq.~[\ref{eq:qed-chemistry-nuclearCBOAHamiltonian}] 
reads as
\begin{align}
\hat{H}_j &=  -\hbar^2\frac{M_1+M_2}{2M_1M_2}\vec{\nabla}_{\textbf{X}}^2 + V_j(\textbf{X})
+\sum\limits_\alpha\frac{1}{2}\left[\hat{p}^2_\alpha+ \omega_\alpha^2\hat{q}^2_\alpha\right] + \omega_\alpha\hat{q}_\alpha {\boldsymbol \lambda_\alpha}  \cdot{\textbf{R}} 
\end{align}
For the numerical calculation we use the following real-space grid: $N_{X}= 61$, $d{X} = 0.08$, $N_{x} = 41$, $d{x} = 0.5$, $N_{\xi}= 51$, $d{\xi} = 0.2$, $N_{pt} = 41$, where the latter describes the maximum amount of photons in the system.

\section*{(SI4) Absorption spectrum}
We calculate absorption spectra with the following formula~\cite{galego2015}
\begin{align}
\sigma(\omega) = \frac{4\pi\hbar\omega}{c}\text{Im} \lim_{\epsilon \rightarrow 0} \sum_k\frac{\left|\bra{\Psi_k}{\textbf{R}}\ket{\Psi_0} \right|^2}{\hbar\omega_k -\hbar\omega_0 - \hbar\omega -\mathrm{i} \epsilon}.
\end{align}
Here ${\ket{\Psi_0}}$ denotes the correlated ground-state with energy $\hbar\omega_0$, while $\ket{\Psi_k}$ are all correlated eigenstates. In the numerical calculations, we apply a broadening of the individual peaks, as presented in Ref.~\cite{flick2014} for electron-phonon problems.

\section*{(SI5) Shin-Metiu model in a cavity}
The Hamiltonian of such a system is given by~\cite{min2014}
\begin{align}
\hat{H}(\textbf{r},\textbf{R}) = &-\frac{\hbar^2}{2M} \vec{\nabla}^2_\textbf{R} -\frac{\hbar^2}{2m_e} \vec{\nabla}^2_\textbf{r} + V_{en}\left(\textbf{r},\textbf{R}\right) +  V_{nn}\left(\textbf{R}\right) +\hat{H}_p
\end{align}  
with the electron-nuclear potential
\begin{align}
V_{en}\left(\textbf{r},\textbf{R}\right) = &V_{en}\left(\left|\textbf{r} - \left(\frac{L}{2},0\right) \right|\right) + V_{en}\left(\left|\textbf{r} - \left(-\frac{L}{2},0\right) \right|\right)+V_{en}\left(\left|\textbf{r} - \textbf{R} \right|\right),
\end{align}
and the nuclear-nuclear potential
\begin{align}
V_{nn}\left(\textbf{R}\right) = &V_{nn}\left(\left|\textbf{R} - \left(\frac{L}{2},0\right) \right|\right) + V_{nn}\left(\left|\textbf{R} - \left(-\frac{L}{2},0\right) \right|\right)+ V_{nn}\left(L\right) + \left(R/R_0\right)^4.
\end{align}
We choose the parameters $a=0.5$, $b=10.0$, $R_0=3.5$, and $L=4\sqrt{3}/5$ as in Ref.~\cite{min2014} and $V_{en}(x) = -1/\sqrt{a + x^2}$ and $V_{nn}(x) = 1/\sqrt{b + x^2}$ with the photon Hamiltonian $\hat{H}_p$ from Eq.~[\ref{eq:ham-photons-appendix1}].

\end{document}